\documentclass[lettersize,journal]{IEEEtran}
\usepackage{amsmath,amsfonts}
\usepackage{algorithmic}
\usepackage{algorithm}
\usepackage{array}
\usepackage[caption=false,font=normalsize,labelfont=sf,textfont=sf]{subfig}
\usepackage{textcomp}
\usepackage{stfloats}
\usepackage{url}
\usepackage{verbatim}
\usepackage{graphicx}
\usepackage{cite}
\usepackage{subfig}
\usepackage{makecell}
\begin{document}

\title{SwinFi: a CSI Compression Method based on Swin Transformer for Wi-Fi Sensing}

\author{Jichen~Bian
\thanks{Jichen~Bian (e-mail: jichen.bian@outlook.com).}
}

\markboth{Journal of \LaTeX\ Class Files,~Vol.~XX, No.~X, XXX~202X}%
{Shell \MakeLowercase{\textit{et al.}}: SwinFi: a CSI Compression Method based on Swin Transformer for Wi-Fi Sensing}

\IEEEpubid{0000--0000/00\$00.00~\copyright~2021 IEEE}

\maketitle

\begin{abstract}
    Wi-Fi sensing is a transformative approach that enables a large of applications through CSI analysis. The challenge lies in the high computational and communication costs with the increasing granularity of CSI data. In this letter, we propose SwinFi, a pioneering solution that compresses CSI at the edge into a succinct feature image and reconstructs at the cloud for further processing. SwinFi employs a Swin Transformer-based autoencoder-decoder architecture that ensures SOTA performance in both CSI reconstruction and sensing tasks. We utilize a dataset for PIR task and conduct extensive experiments to evaluate SwinFi. The results show that SwinFi achieves the reconstruction quality with the NMSE of -37.74dB and the classification accuracy of 95.3\% at the same time. 
\end{abstract}


\section{Introduction}
\IEEEPARstart{W}{i-Fi} sensing has emerged as a pivotal technology in a variety of applications, ranging from human activity recognition to object sensing and localization. Its widespread adoption is underpinned by the unique advantages it offers over traditional sensing methods, notably its cost-effectiveness, device-free nature, and privacy-preserving characteristics. This burgeoning field leverages the extraction of Channel State Information (CSI) to establish key mappings from channel states to recognition targets, facilitating a multitude of sensing capabilities \cite{10.1145/3310194}. The advancement of Multiple-Input Multiple-Output (MIMO) and Orthogonal Frequency-Division Multiplexing (OFDM) technologies endow CSI with higher granularity, providing robust support for Wi-Fi sensing technologies. This enhancement, however, brings a substantial increase in computational resource requirements. While cloud and edge computing capabilities offer viable solutions to alleviate computational burdens, they concurrently impose a considerable strain on communication systems due to the extensive data transmission involved. In response, CSI compression technology stands out for its ability to markedly reduce data processing and transmission requirements, thereby bolstering the efficiency and practicality of Wi-Fi sensing systems.

Recent developments in deep learning contribute to new methods in the field of CSI compression. \cite{8322184} introduces CsiNet, which applies convolutional neural networks (CNN) for the learning and reconstruction of CSI features. Building on this, further studies incorporated temporal dynamics into CSI processing. The integration of CNN with Long Short-Term Memory (LSTM) networks in \cite{8482358} enables the extraction of both spatial and temporal features from CSI data, leading to enhancements in reconstruction accuracy, particularly in environments with dynamic network conditions. Additionally, \cite{9718553} explores the application of Transformer models, utilizing multi-head self-attention (MSA) mechanisms to adeptly handle long-range dependencies in CSI data. 

However, the aforementioned works are primarily oriented towards 5G communication, where CSI feedback plays a crucial role in ensuring high-quality precoding and efficient communication transmission, with the emphasis on correct reconstruction. \cite{9667414} presents a different perspective, asserting that CSI compression for Wi-Fi sensing fundamentally diverges from the approach in 5G communication. Although the tasks in both domains are similar, in Wi-Fi sensing, the compressed CSI data requires not only reconstructability but also discriminative feature to support various sensing applications. In other words, the compressed feature map should preserve essential characteristics for both recognition and sensing tasks.

In this letter, we propose an innovative method for CSI compression of Wi-Fi sensing. Inspired by \cite{9667414}, we focus on both reconstruction accuracy and classification features as key metrics. To capture the spatial and temporal dynamics of CSI, we treat a sequence of CSI matrices over time as a unique form image. Leveraging the principles of Swin Transformer, we design an autoencoder-decoder architecture tailored for Wi-Fi sensing applications. The contributions of this letter are summarized as follows:

\IEEEpubidadjcol

\begin{enumerate}
    \item To reduce the communication and computational costs of Wi-Fi sensing, we propose SwinFi, a novel method that compresses CSI data into a compact feature image at the edge and reconstructs the original CSI data at the cloud for further processing.
    \item We design a multitask joint model that, building upon the SwinFi Encoder, integrates a linear classification head to enable the recognition of categorical features within CSI feature images.
    \item To evaluate the performance of SwinFi, we produce a dataset for personnel identity recognition (PIR) tasks based on Wi-Fi CSI. We compare SwinFi with several state-of-the-art (SOTA) methods and the results show that SwinFi achieves the best reconstruction quality with the Normalized Mean Squared Error (NMSE) of -37.74dB, while concurrently attaining a classification accuracy of 95.3\%.
\end{enumerate}

\IEEEpubidadjcol

\section{Methodologies}

We tend to adapt the latest computer vision (CV) methods to CSI processing. Therefore, in this section, we first introduce two SOTA methods based on Transformer. Then we delve into the architecture and details of our proposed SwinFi.

\subsection{Vision Transformer}

Vision Transformer (ViT) is introduced in \cite{dosovitskiy2021image}, representing a significant breakthrough in CV. ViT adapts the Transformer architecture, traditionally used in natural language processing (NLP), for image analysis tasks. This adaptation demonstrates remarkable success in CV, including classification, segmentation, and object detection.

The core idea of ViT is to treat an image as a sequence of fixed-size patches, analogous to the words in a text. The patches are then linearly embedded into a sequence of vectors. The model applies a Transformer encoder to these embeddings to capture the complex relationships between the patches. The encoder consists of multiple layers, each containing two main components: MSA and a fully connected feed-forward network. After the encoder, the output is leveraged for various downstream tasks.

\vspace{-10pt}
\subsection{Swin Transformer}

While ViT treats images as sequences of patches similar to words, this approach leads to exponential increases in computational cost with image size due to global MSA. Furthermore, the patch method may overlook the details within itself, leading to a loss of local structural information.

Based on ViT, Swin Transformer \cite{9710580} proposes several key innovations that further adapt the Transformer architecture for CV. Swin Transformer incorporates the notion of local windows into ViT architecture, similar in spirit to the receptive fields of convolutional neural networks (CNN). The input image is divided into patches smaller than ViT does. Instead of processing all patches at once, Swin Transformer applies MSA within local windows. Another pivotal innovation is the shift window mechanism, which enables information to flow across the entire image by alternating the window positions across layers. This mechanism effectively achieves a global receptive field while maintaining a linear computational complexity with the image size. The complexity comparison between MSA and window MSA (W-MSA) is as follows: 
\begin{equation}
    \label{swinomega}
    \begin{aligned}
        \Omega(MSA) = 4hwC^2 + 2(hw)^2C, \\
        \Omega(W-MSA) = 4hwC^2 + 2M^2hwC,
    \end{aligned}
\end{equation}
where \(h\) and \(w\) are the patch number, \(C\) is the dimension of patch, and \(M\) is the window size.

\vspace{-10pt}
\subsection{SwinFi}

In the context of CSI compression and reconstruction, the data, characterized by its multi-dimensional nature (subcarriers, time, and multiple antennas), presents a formidable challenge. To adapt CV methods for CSI data, we consider each CSI matrix as an "image" where each pixel represents the amplitude or phase of a subcarrier at a specific time. 

\begin{figure*}[htbp] 
\centerline{\includegraphics[width=1.0\linewidth]{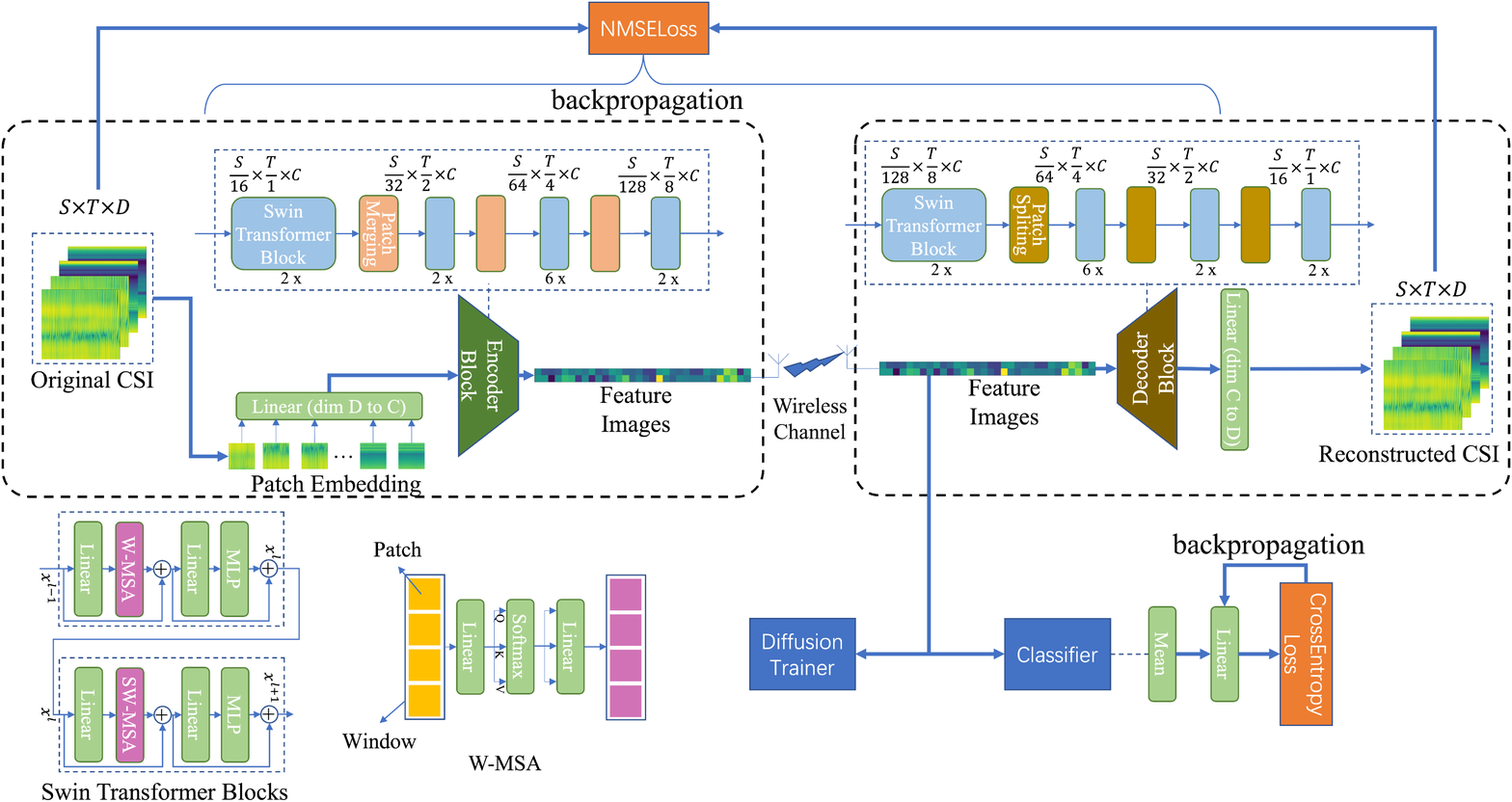}} 
\caption{The architecture of SwinFi.} 
\label{model}
\end{figure*}

Inspired by Swin Transformer, we propose SwinFi as shown in Fig~\ref{model}, a novel method employing an end-to-end autoencoder-decoder architecture, specifically for CSI compression and PIR tasks. SwinFi is designed to capture the spatial and temporal dynamics of CSI data, enabling efficient and effective compression. The components of SwinFi are as follows:

1. \textbf{Encoder}: Encoder consists of a patch embedding block and an encoder block. The patch embedding block transforms the input \(\mathbf{X} \in \mathbb{R}^{B \times D \times S \times T}\) into \(\mathbf{X} \in \mathbb{R}^{B \times N \times C}\), where \(B\) is the batch size, \(D\) is the number of channels, \(S\) is the number of subcarriers, \(T\) is the number of time slots, \(N\) is the number of patches, and \(C\) is the patch embedding dimension. The patch focuses on the temporal dimension, with each patch being of size \(p_S \times p_T \times D\). The encoder block comprises several Swin Transformer Block layers and Patch Merge layers, each Swin Transformer Block including a W-MSA, a window position encoding layer, and an MLP layer.

2. \textbf{Decoder}: Decoder mirrors the architecture of Encoder but with Patch Merge layers replaced by Patch Split layers. It decodes the encoded CSI feature image back into the original space. During training, the parameters of Encoder and Decoder are jointly optimized to minimize the NMSELoss between the original CSI data and the reconstructed data.

3. \textbf{Classifier}: Classifier is a straightforward linear structure to identify the categorical features within feature images for PIR tasks. Training of Classifier utilizes the CrossEntropyLoss to update its parameters.

It is noteworthy that, unlike the square shapes commonly utilized in CV, both the patch and window sizes in SwinFi adopt rectangular dimensions, with a patch width of 1, meaning that each time slot occupies a separate patch. Similarly, the windows have a length of 1, as we aim to avoid computing MSA across patches that are neither adjacent in time slots nor subcarriers, a measure grounded in the belief that such quantities lack physical significance. The superiority of this approach will be substantiated in the experimental section.

\section{Experimental Results}

In this section, we first present the CSI datasets made for PIR tasks. Then we introduce the experimental setup and evaluation of SwinFi.

\subsection{Datasets}

\begin{figure}[htbp] 
\centerline{\includegraphics[width=0.8\linewidth]{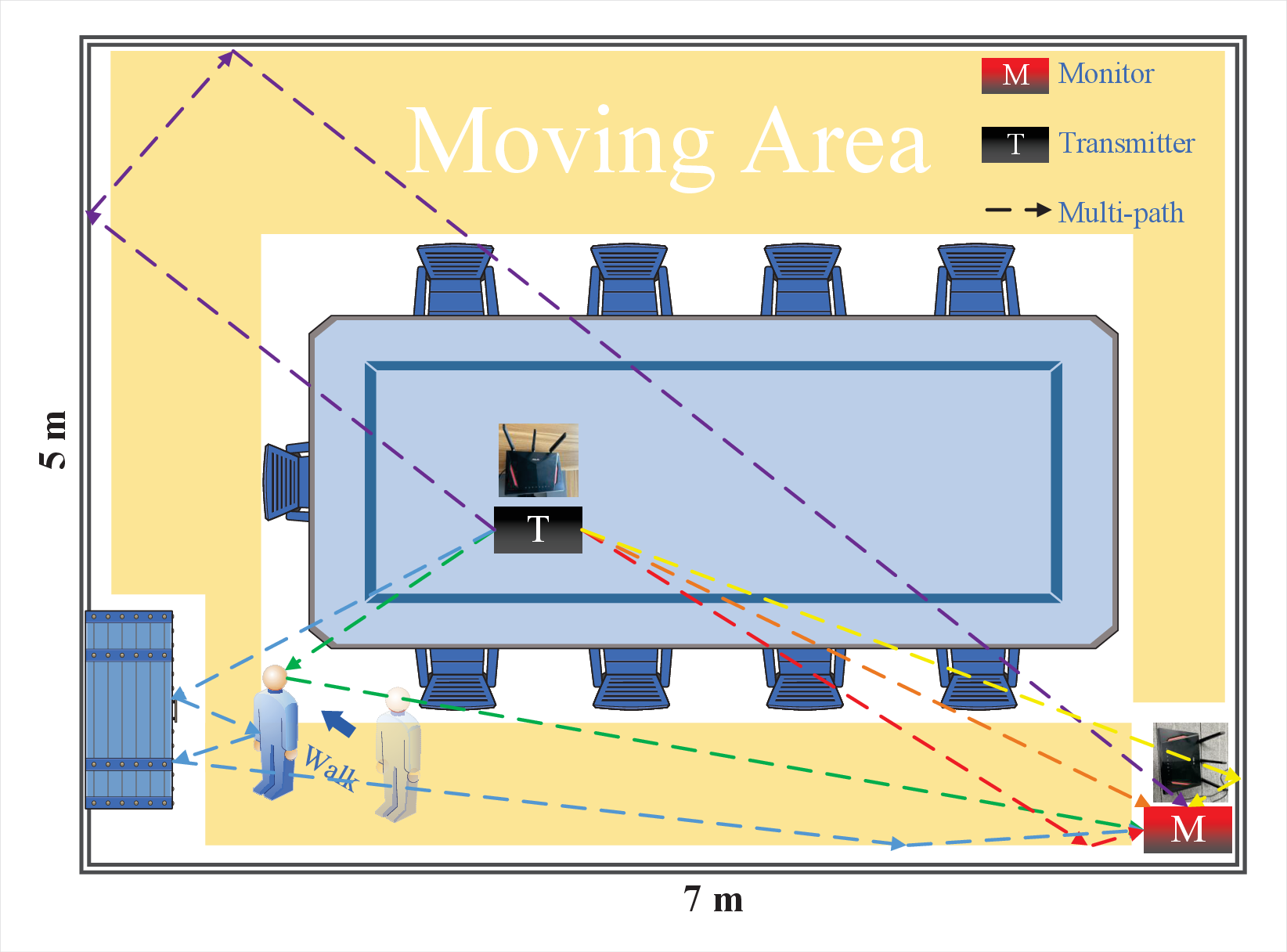}} 
\caption{Layout of the experimental scene.} 
\label{meetingroom}
\end{figure}

In this letter, we produce a specialized Wi-Fi CSI dataset designed for PIR tasks. The dataset is assembled using two 802.11ac routers equipped with Nexmon tools \cite{10144501, 10.1145/3349623.3355477}, set up in a meeting room as depicted in Figure~\ref{meetingroom}. We recorded the CSI data of 20 participants, each freely moving within the room. A single-antenna transmitter transmits signals at an 80MHz frequency every 10 milliseconds, while a receiver with four antennas captures the CSI data, which includes gait characteristics crucial for identifying individuals.

The raw communication cost of this setup, without any preprocessing, is calculated to be \(4 \times 256 \times 100 \times 2 \times 4\) Bytes/s, equivalent to approximately 6.55Mbps. This corresponds to \(8\) bytes per measurement, with each complex number (detailing amplitude or phase) stored using two floating points: 4 bytes each for the real and imaginary parts. Based on the aforementioned setup, we conduct experiments that involve collecting five minutes of CSI data for each of the 20 participants, as well as for an empty meeting room, resulting in a dataset that categorizes 21 different classes.

As identified in \cite{8259000}, phase errors in CSI measurements can stem from several sources related to hardware imperfections. These include packet detection delay (PDD), sampling frequency offset (SFO), carrier frequency offset (CFO), random initial phase offset, and phase ambiguity. Such errors compromise the accuracy of phase information.

\begin{figure}[htbp]
\centering
{\includegraphics[width=0.30\linewidth]{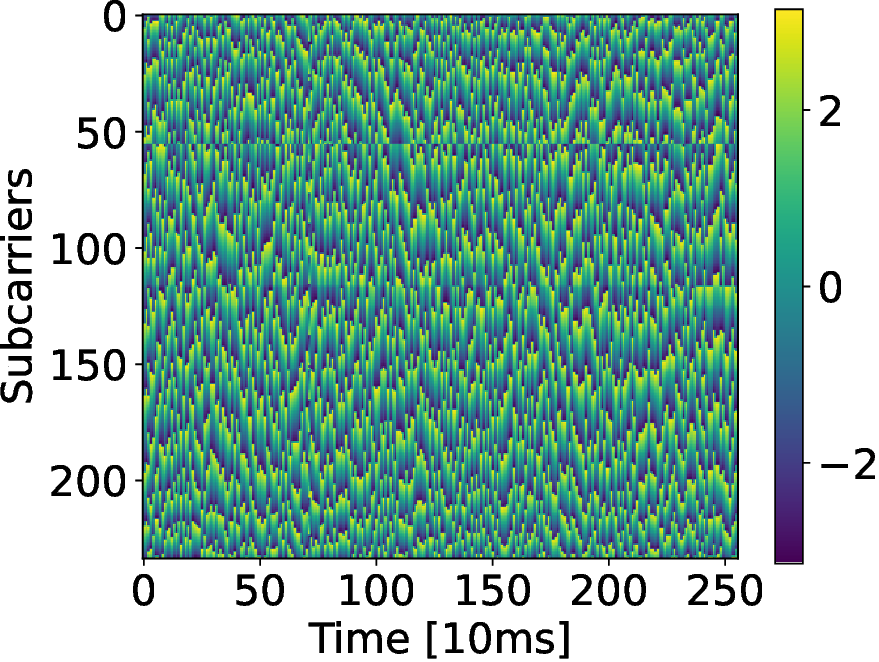}}
{\includegraphics[width=0.32\linewidth]{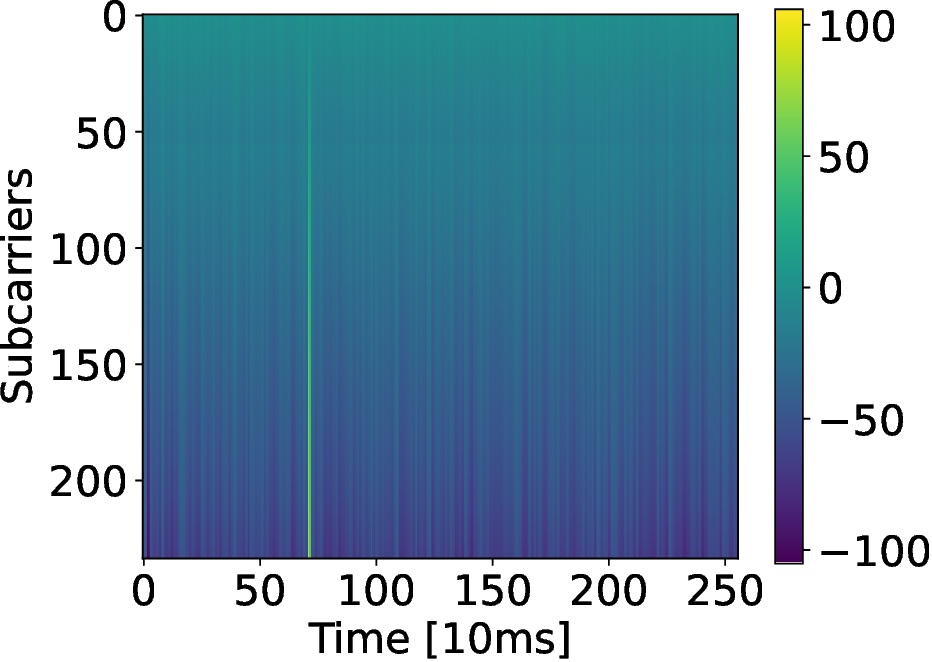}}
{\includegraphics[width=0.30\linewidth]{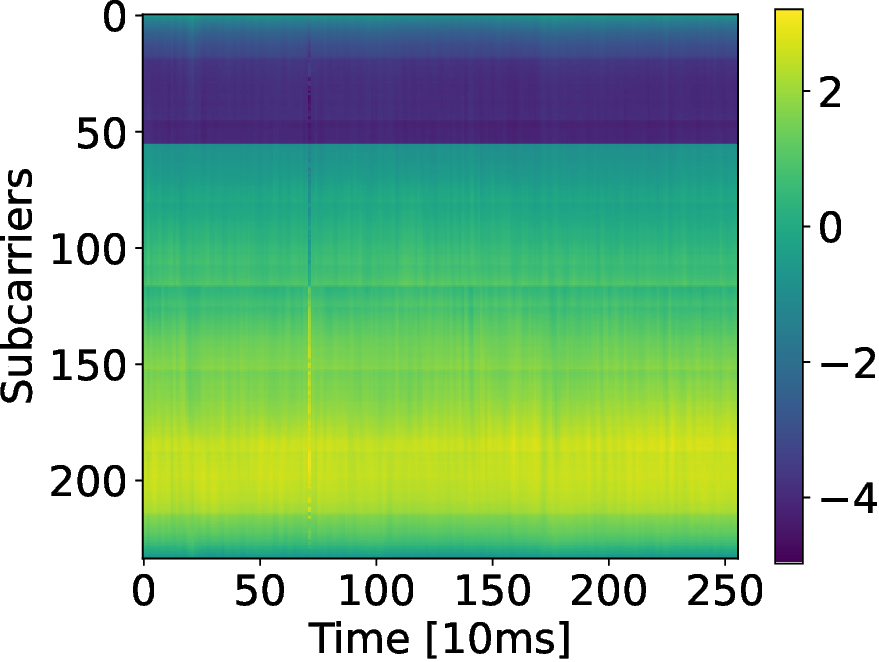}}
\caption{Phase error processing, illustrated in three stages. On the left, raw phase data are presented, showcasing inherent noise and discontinuities. The figure in the center depicts the phase after unwrapping, where discontinuities are corrected. Finally, the figure on the right demonstrates the phase error after processing, representing the linear fitting to rectify the phase error.}
\label{phase-error}
\end{figure}

To address these issues, we implement a two-step correction process, as shown in Fig~\ref{phase-error}. The initial step involves phase unwrapping along the subcarrier dimension, effectively mitigating discontinuities that arise due to the cyclic nature of phase measurements. This unwrapping process is described as:

\begin{equation}
    \hat{\phi}_{m+1} = \begin{cases} 
    \hat{\phi}_{m+1} - 2\pi & \text{if } \hat{\phi}_{m+1} - \hat{\phi}_m \geq \pi, \\
    \hat{\phi}_{m+1} + 2\pi & \text{if } \hat{\phi}_{m+1} - \hat{\phi}_m \leq -\pi, \\
    \hat{\phi}_{m+1} & \text{otherwise},
    \end{cases}
    \label{unwrapped}
\end{equation}

where $\hat{\phi}_k$ represents the phase of the $k$-th subcarrier.

Following the unwrapping, the phase data is subjected to a linear fitting process. The process is governed as:

\vspace{-10pt}
\begin{equation}
    a = \frac{\hat{\phi}_n - \hat{\phi}_1}{k_n - k_1} - \frac{2\pi\delta}{N},
\end{equation}

\vspace{-5pt}
\begin{equation}
    b = \frac{1}{n} \sum_{j=1}^{n} \hat{\phi}_j - \frac{2\pi\delta}{nN} \sum_{j=1}^{n} k_j + \beta,    
\end{equation}

\vspace{-5pt}
\begin{equation}
    \tilde{\phi}_j = \hat{\phi}_j - (ak_j + b),    
\end{equation}

where \(k_j\) denotes the index of the \(j\)-th subcarrier, and \(a\) and \(b\) are the coefficients determined through the fitting.
The real phase is then recalculated by subtracting the derived linear term \(ak_j+b\) from the measured phase \(\hat{\phi}_j\), resulting in a corrected phase profile. 

\begin{figure}[htbp]
\centering
\subfloat[Amp (participant 4)]
{\includegraphics[width=0.45\linewidth]{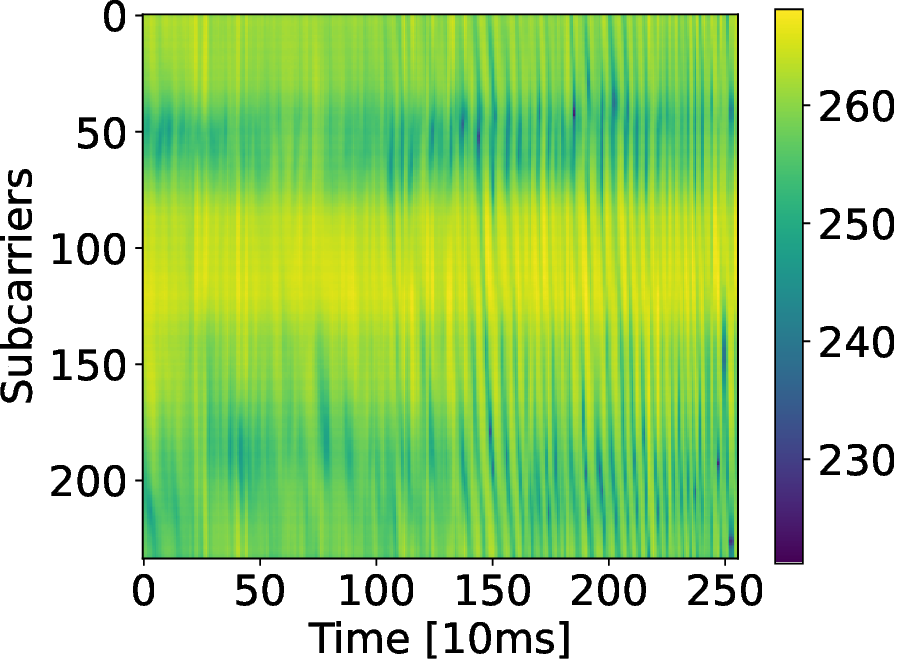}\label{amp4}}
\subfloat[Phase (participant 4)]
{\includegraphics[width=0.45\linewidth]{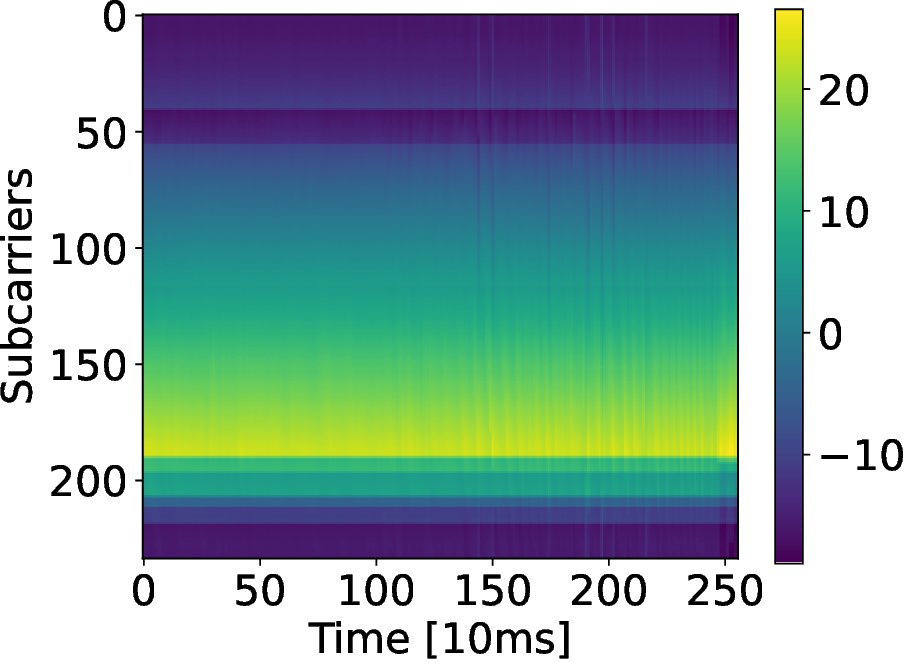}\label{pha4}}\\
\vspace{-10pt}
\subfloat[Amp (participant 5)]
{\includegraphics[width=0.45\linewidth]{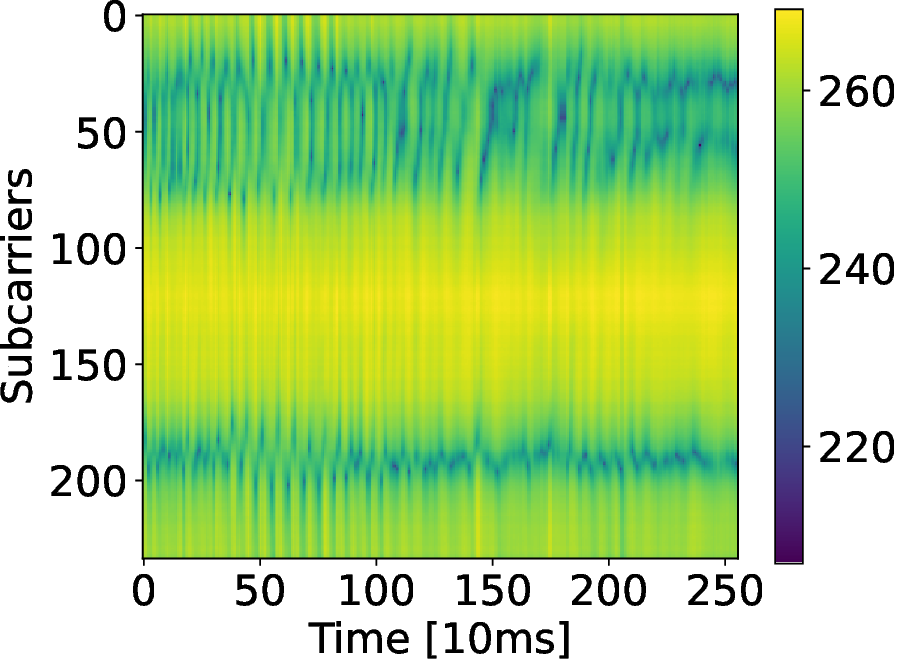}\label{amp5}}
\subfloat[Phase (participant 5)]
{\includegraphics[width=0.45\linewidth]{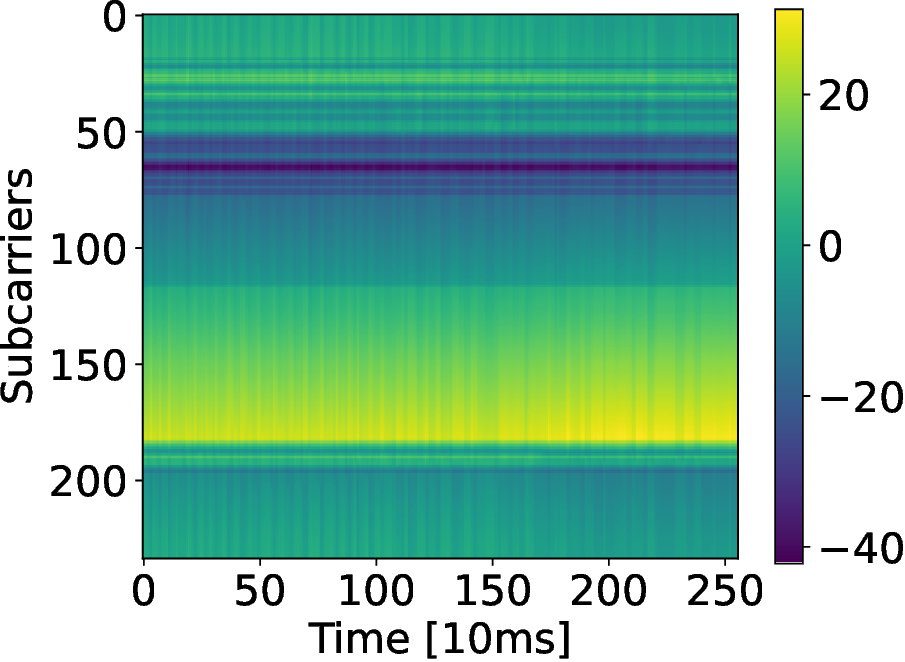}\label{pha5}}
\caption{Amplitude (in dB scale) and phase of CSI data for the participant 4 (a)-(b) and 5 (c)-(d). Each trace is about 2.56 seconds long (\(x\)-axis) and shows the amplitude or the phase on each subcarrier (\(y\)-axis). Note that CSI is not available on the control carriers and pilot carriers, so there are 234 carriers left to use.}
\label{csi-img}
\end{figure}

Here are the examples of CSI data, as shown in Fig~\ref{csi-img}.

\subsection{Experimental Setup}

\textit{Criterion:} For evaluating the performance of models, the primary factors we consider are the rate of compression, the quality of reconstruction, and the accuracy of classification. The compression rate is calculated as the ratio of the communication cost of the raw data to the communication cost of the compressed data, as follows:

\begin{equation}\label{gamma}
\begin{aligned}
    \gamma &= \frac{S \times T \times D}{\frac{S}{p_S} \times \frac{T}{p_T} \times C / 4^{len(depth)-1}} \\
    &= \frac{p_S \times p_T \times D \times 4^{len(depth)-1}}{C},
\end{aligned}
\end{equation}
where \(C\) is the number of feature dimensions, \(p_S\) and \(p_T\) are the patch sizes in the spatial and temporal dimensions, respectively.

The quality of reconstruction is measured by NMSE in decibels (dB), calculated as:

\begin{equation}
    \text{NMSE} = 10 \times \log_{10}\left(\frac{\sum_{i=1}^{N}||\mathbf{X}_i - \hat{\mathbf{X}}_i||^2}{\sum_{i=1}^{N}||\mathbf{X}_i||^2}\right),
\end{equation}

The classification accuracy is calculated as the percentage of correctly classified samples.

\textit{Baseline:} We compare the performance of SwinFi with several SOTA methods. For the single task of CSI reconstruction, we compare with CSINet \cite{8322184}, WiWho \cite{7460727}, and AutoID \cite{10.5555/3504035.3504244}. For the task of PIR, we compare with SimpleViTFi \cite{SimpleViTFi}. For the joint task, the relevant method is EfficientFi \cite{9667414}.

\subsection{Evaluation and Disscussion}

We first evaluate the performance of SwinFi with the baseline methods mentioned above in Table~\ref{results-baseline}. From Equation~\ref{gamma}, the compression rate \(\gamma\) is decided by several factors. Given our intention to maintain a finely-grained window receptive field through the use of small patch sizes, we are thus constrained to modulate the compression ratio \(gamma\) by adjusting the feature dimensions and the configuration of Swin Transformer blocks. We design a series of experiments as shown in Table~\ref{experiments-setup}, allowing the compression ratio to vary from 64 to 1024. Note that the mixed data contain amplitude and phase for each of the 4 channels. To maintain the same compression ratio, \textbf{Dim} is doubled.

\begin{table}[htbp]
    \centering
    \caption{Setup of Different Compression Ratio}
    \begin{tabular}{ccccc}
    \hline
    \textbf{Input} & \textbf{Patch\_size} & \textbf{Dim} & \textbf{Depth} & \textbf{$\gamma$}\\ \hline
    Amplitude or & $8 \times 1$ & 32 & [2,2,6,2] & 64 \\ 
    Phase & $8 \times 1$ & 16 & [2,2,6,2] & 128 \\ 
    $4 \times 256 \times 256$ & $8 \times 1$ & 32 & [2,2,2,6,2] & 256 \\ 
        & $8 \times 1$ & 16 & [2,2,2,6,2] & 512 \\ 
        & $8 \times 1$ & 32 & [2,2,2,2,6,2] & 1024 \\ \hline
    Mixed & $8 \times 1$ & 64 & [2,2,6,2] & 64 \\ 
    $8 \times 256 \times 256$ & $8 \times 1$ & 32 & [2,2,6,2] & 128 \\ 
        & $8 \times 1$ & 64 & [2,2,2,6,2] & 256 \\ 
        & $8 \times 1$ & 32 & [2,2,2,6,2] & 512 \\ 
        & $8 \times 1$ & 64 & [2,2,2,2,6,2] & 1024 \\ \hline 
    \end{tabular}
    \label{experiments-setup}
\end{table}

In Table~\ref{results-baseline}, it is observed that SwinFi outperforms the baseline methods in terms of the least reconstruction error with the NMSE of -37.74dB. Even at high compression rates, SwinFi maintains a high quality of reconstruction along with superior classification accuracy. This performance stands on par with, or even surpasses, that of single-task models.

\begin{table}[htbp]
    \centering
    \caption{Comparison of different methods.}
    \begin{tabular}{l|ccc}
    \hline
    \textbf{Method} & \textbf{$\gamma$} & \textbf{NMSE (dB)} & \textbf{Accuracy (\%)} \\ \hline
    CSINet \cite{8322184} & 4 & -29.18 & N/A \\
        & 16 & -26.18 & N/A \\
        & 32 & -20.40 & N/A \\
        & 64 & -18.07 & N/A \\ \hline
    WiWho \cite{7460727} & N/A & N/A & 67.3 \\ \hline
    AutoID \cite{10.5555/3504035.3504244} & N/A & N/A & 77.6 \\ \hline
    SimpleViTFi \cite{SimpleViTFi} & N/A & N/A & 96.7 \\ \hline
    EfficientFi \cite{9667414} & 66.8 & -35.18 & 84.5 \\
        & 148.4 & -34.23 & 81.6 \\
        & 334.0 & -30.19 & 82.7 \\
        & 763.4 & -29.18 & 82.1 \\
        & 1781.3 & -27.70 & 83.3 \\ \hline
    \textbf{SwinFi} & 64 & -37.74 & 95.3 \\
    & 128 & -37.56 & 93.8 \\
    & 256 & -35.81 & 92.2 \\
    & 512 & -31.47 & 90.4 \\
    & 1024 & -29.19 & 87.3 \\ \hline
    \end{tabular}
    \label{results-baseline}
\end{table}

The baseline methods focus exclusively on CSI amplitude. Accordingly, the comparative experiments of SwinFi also utilize only the amplitude. Subsequently, we conduct a series of experiments on both the phase and mixed data of CSI. These experiments aim to provide a comprehensive understanding of the extent to which SwinFi can effectively process and leverage the full spectrum of CSI signal components.

Table~\ref{results-amppha} provides a comparison of the performance across phase and mixed data. Combined with amplitude results in Table~\ref{results-baseline}, while SwinFi can handle phase data effectively, the reconstruction quality and classification accuracy are less satisfactory. However, by mixed data, SwinFi achieves the best classification accuracy. This suggests that the distinct features of human gait, used for classification, are distributed across both amplitude and phase components of the CSI data.

\begin{table}[htbp]
    \centering
    \caption{Comparison of different inputs.}
    \begin{tabular}{l|ccc}
    \hline
    \textbf{Input} & \textbf{$\gamma$} & \textbf{NMSE (dB)} & \textbf{Accuracy (\%)} \\ \hline
    Phase & 64 & -15.16 & 93.3\\
        & 128 & -15.73 & 80.2 \\
        & 256 & -16.85 & 91.4\\
        & 512 & -16.14 & 84.4\\
        & 1024 & -13.87 & 89.4\\ \hline
    Mixed & 64 & -30.13 & 97.8 \\
        & 128 & -27.08 & 93.8 \\
        & 256 & -25.24 & 98.3 \\
        & 512 & -25.13 & 93.4 \\
        & 1024 & -22.93 & 89.6 \\ \hline
    \end{tabular}
    \label{results-amppha}
\end{table}

Finally, we investigate the impact of different patch shapes and window shapes on the performance of SwinFi. The results are shown in Table~\ref{results-window}. It is observed that the rectangular shapes outperform the square shapes in both NMSE and classification accuracy under the similar $\gamma$. This result is consistent with the intuition that rectangular shapes can better capture the spatial and temporal dynamics of CSI data.

\begin{table}[htbp]
    \centering
    \caption{Comparison of different patch size and window size.}
    \begin{tabular}{cc|ccc}
    \hline
    \textbf{Window\_size} & \textbf{Patch\_size} & \textbf{$\gamma$} & \textbf{NMSE (dB)} & \textbf{Accuracy (\%)} \\ \hline
    $4 \times 4$ & $3 \times 3$ &  72 & -32.63 & 90.8 \\
    $4 \times 4$ & $3 \times 3$ & 288 & -27.79 & 89.2 \\ \hline
    $1 \times 16$ & $8 \times 1$ & 64 & -37.74 & 95.3 \\
    $1 \times 16$ & $8 \times 1$ & 256 & -35.81 & 92.2 \\ \hline
    \end{tabular}
    \label{results-window}
\end{table}

\section{Conclusion}

In this letter, we introduce SwinFi, leveraging Swin Transformer within an autoencoder-decoder architecture for the efficient compression and reconstruction of CSI data in Wi-Fi sensing. The approach not only reduces computational and communication demands but also demonstrates SOTA performance in reconstruction quality and classification accuracy. Through extensive experiments, SwinFi achieves satisfactory results and outperforms existing methods. 

Anticipating future research, we will explore the integration of compressed feature images with Diffusion model as shown in Fig~\ref{model}. Our aim is to investigate the potential of Diffusion model for generating synthetic CSI data to augment the existing datasets. Considering computational efficiency, we propose utilizing the compressed latent variables as inputs to Diffusion, where the output generated data are then reconstructed to achieve a lightweight generative process.

\bibliographystyle{IEEEtran}
\bibliography{reference.bib}

\end{document}